# A Three-Dimensional GUI for Windows Explorer

David Carter
Luiz Fernando Capretz
University of Western Ontario
Dept. of Electrical & Computer Engineering
London, Ontario, N6G 1H1, CANADA
{dcarter2, lcapretz}@uwo.ca

## Abstract

*Three-dimension will be a characteristic of future user interfaces, although we are just starting to gain an understanding of how users can navigate and share information within a virtual 3D environment. Three-dimensional graphical user interfaces (3D-GUI) raise many issues of design, metaphor and usability. This research is devoted to designing a 3D-GUI as a front-end tool for a file management system, in this case, for Microsoft Windows© Explorer; as well as evaluating the efficiency of a 3D application. The software design was implemented by extending the Half-Life 3D engine. This extension provides a directory traversal and basic file management functions, like "cut", "copy", "paste", "delete", and so on. This paper shows the design and implementation of a real-world application that contains an efficient 3D-GUI.*

## 1 Introduction

The graphical user interface (GUI) was developed at Xerox PARC in the late seventies. It was successfully commercialised by Apple with the Macintosh computer in the early eighties and has since become the integral part of every modern operating system for personal computers and graphics workstations. One reason for this growth is productivity: a number of studies have shown GUIs, with their direct manipulation style of interaction, enhance productivity [1]. Another reason is subjective preference: people express a preference for GUI interfaces, which is now recognized as a key, often taken for granted, feature of any software systems.

Computer users are currently bound to the "desktop" metaphor. This metaphor's longevity is a testament to the strength of its design. Of course, over twenty-some years improvements have been made to that original design, but the basic elements (e.g. icons, pop-up and pull-down menus) that implement the *what-you-see-is-what-you-get* idea have remained the same. Nevertheless, our everyday life is 3D, therefore a natural interface would be a three dimensional one.

To gain perspective of the current situation it is necessary to look at the properties of the desktop metaphor. The user interface exists in the xy-plane, thus giving it two dimensions (and four degrees of freedom). Because of the overlapping of several independent windows, a pseudo z-plane is defined through the obscuring of "background" windows. Thus a full third dimension does not exist, but it is not limited to two dimensions either. It is somewhere in between; in fact, this has become a pseudo-2.5D user interface [2].

Because of the increase in development of 3D hardware and software, the customer base in terms of personal computing has gone from scientists to the ordinary PC user, with interface application spanning from medicine to CAD/CAM engineering. Thus the 3D user interface no longer concerns only the most knowledgeable people but rather the average home user, who should, without qualified assistance or even without referencing the user manual, be able to start an application within the 3D environment with minimal knowledge of 3D computing [3].

Just as decreasing hardware price and increasing hardware capabilities made the pseudo-2.5D GUI affordable in the eighties and widespread in the nineties, these hardware trends will make three-dimensional GUIs affordable. GUI with 3D capability offers great potential for improvement over today's 2.5D GUIs. 3D graphics look nice and are part of nearly every consumer PC now, and it will become even more pervasive in the near future [4].

With the current situation in mind, it is now possible to look at the theory behind 3D graphical user interfaces. The 3D space consists of the xyz-space in which elements are situated. Since it employs all three planes (xy, xz, yz) it is a genuine three-dimensional representation. However, unlike the desktop metaphor, design considerations are far more complex. To start, there is no standard metaphor that defines the organization of objects in a 3D space.

Since no one metaphor is dominant, why even attempt to use it? The answer to this is simple. Metaphors provide the end user with a familiar setting in an unfamiliar environment. Tailoring the environment to meet the needs of the user by giving them a sense of familiarity is intended to increase the user's productivity.

Within the realm of 3D user interfaces there are many different metaphors being considered and implemented. The first of these are desktop extensions, such as Data Mountain [5], which has begun to explore 3D visualization for every day office work. Also included in the desktop extension is the hallway metaphor; in this environment, 2D windows are hung on the walls of a 3D hallway for the user to view as he/she walks by. Leach et al. [2] presents a metaphor used for a 3D GUI in which windows are arranged in a *tunnel*. The user is positioned in the middle of the mouth of the *tunnel* looking toward the other end. Windows are displayed with a perspective projection. In addition to the front-end window, there is a "hanging" mode where the windows are hung on the left or right wall of the tunnel. Another new idea is a 3D cursor with six degrees of freedom, called the *magic wand*, which "floats" over the top of objects rather than being part of the screen.

Although games are still the primary PC-based 3D users, any user would also benefit from seeing their concepts in 3D. This is precisely the idea behind ROOMS [6]. This gives the user the illusion of a 3D desktop by displaying 3D wallpapers in the background of the environment, just like computer games. 3D desktop wallpapers allow the user to move icons into and manipulate objects as in a 3D game.

However, within a single metaphor it is still very possible for the user to become disoriented and "lost" because of the user's unfamiliarity with the 3D experience itself. Users tend to have a difficult time controlling and manipulating objects within a 3D space when they cannot employ their hands. To aid in the problem of becoming disoriented and lost, a meta-view of the 3D can be considered to show the users either where they are or what they are looking at.

The purpose of this research was to develop and evaluate a three-dimensional user interface as a front end for a file management system (such as Microsoft Windows© Explorer) using a meta-view approach. The implemented software, fully described in [7], reused a modified version of the user interface used in Valve Software's Half-Life engine.

## 2 File Manager Requirements

Although it seems reasonable that the home users should be able to fully immerse themselves in a 3D rendered world, it is not always possible to efficiently provide this. Typically, 3D interfaces are often awkward for the developers to represent data, and harder for the users to manipulate than the standard 2D user interface.

In considering 3D graphical user interfaces, it became evident that if an effective 3D user interface were to be developed it needed to be the front-end for a practical application. Thus the idea was to provide the user with a system for managing files, one that incorporated 3D rendered graphics, yet at the same time maintained the same (if not a higher) level of usability as the standard 2D interface to which the home user had become accustomed.

The developed software was intended to be an extension of the standard Microsoft Windows© Explorer. It enables the user to traverse a file system from a first person perspective using a 3D interface, and it provides the functionalities listed in Table 1.

Table 1: Functionalities of the 3D user interface.

| File Manager | User Interface |
| --- | --- |
| Create a New Directory | Generate the interface |
| Rename | Inventory |
| Cut | Return to Root |
| Copy | Map Layover |
| Paste | Select All |
| Delete | Refresh |

The file manager functionalities reflect the most common tasks performed by the average user. The user interface functionalities were created to aid the user in carrying out the file management services.

## 2.1 The User Profile

The software is intended for two different types of users. The first type is the person unfamiliar with the file structure employed by an operating system and who also has trouble visualizing the file system in a two dimensional manner. For him/her, the software is useful since it creates a structured environment that is analogous to a file system and also it allows for a visual learning process to commence. The second type of user is the one familiar with first person perspective employed in many games; for him/her, the software becomes an interactive means of monitoring the user's file system.

## 2.2 The User Interface

The software implements a modified version of the Half-Life engine and is employed to represent the directories and files contained within the file system.

Directories are represented by rectangular rooms. Along two of the four walls in the room transporters to the subdirectories are placed, and along one of the remaining walls a transporter to the parent directory is also placed.

Files were originally intended to be represented as their native Microsoft Windows© icons; however, as development progressed, the need for a more reusable representation arose, and files are represented as panes of glass hovering in their appropriate room.

## 3 The Design of 3D File Manager

Originally it was thought that the software would consist of three main components: the file manager, the 3D user interface, and the Half-Life engine. However, as the design was refined portions of these components shifted, a more efficient design emerged.

Since the Half-Life engine follows the client-s architecture, it was able to absorb the other two components. The file management component was consolidated with the server side, while the client side absorbed the user interface functions. However, it was also necessary to develop software to generate the data needed for the Half-Life engine to represent the file system. Thus the system contains three components.

As depicted in Figure 1, the *InitLevel* class is responsible for generating the data needed for the Half-Life engine. The *Half-Life Engine* package contains both the server and client components. The *DirectoryInfo* and *FileInfo* classes are taken from the .NET framework and are used to aid in the file system traversal.

In using the Half-Life engine, it was possible, by reusing the necessary portions of the engine and then adding the file management and user interface functions, to produce the desired 3D user interface. However, in using the Half-Life engine, the major stumbling block became the compatibility issues between Microsoft Windows© and the engine (an OpenGL based system).

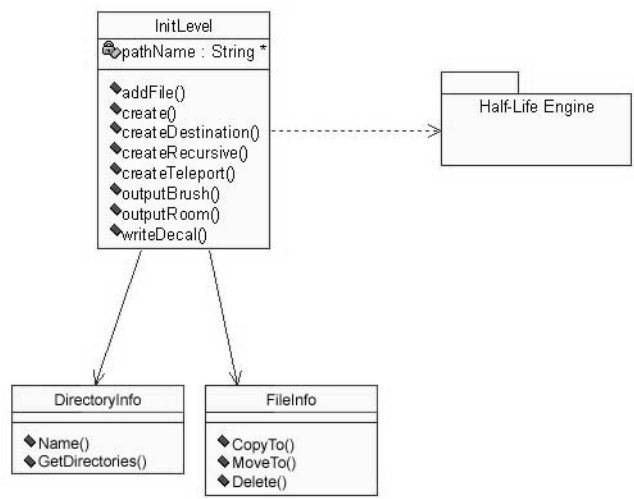

Figure 1: Class diagram for the 3D file manager.

## 4 Implementation

The implementation of the 3D file management system took place in two phases. First, the map generator (implemented in Visual C++ .NET) was developed to build the virtual world the user could walk through. Secondly, to add the file management capabilities promised in the requirements, alterations were made to the Half-Life engine.

In originally creating the design for the 3D file management system, it was not clearly understood what was and was not possible with the available development software. That is, a map was generated by the system after recursively searching the directories within it. That map was compiled and passed off to the Half-Life engine, where it was rendered so that the user could traverse it and perform the specified operations.

However, the details of each of these functions altered greatly as three concepts became clear:

First was the ease with which the .NET Framework was able to traverse the file system to provide the basis for the maps.

Secondly was the complexity of generating the map files. Even though before they are compiled the files are simply plain text, the difficulty in creating these maps, in terms of their size and the order in which each of the brushes or entities must be added became increasingly apparent.

Thirdly was that the sheer size of the Half-Life engine became overwhelming. The complexity of the source code, in terms of how the various modules of the engine communicate, made understanding it and adapting it rather difficult.

In order to adapt to the changing climate of the project, it was necessary to alter - and in some cases overhaul - the design that was originally created. However, in doing so, the opportunity for constant inspections and revisions were created.

## 5 Testing

Performance is a major issue in any interface. We found that the 3D file manager performs satisfactorily on a Pentium-based personal computer (without a 3D graphics accelerator board) running Windows 2000.

Although there are many existing benchmarks for the Half-Life engine, it is necessary to test the performance as it relates to the rendering of the generated map, since the entities, brushes, textures, and the design of the map itself have an effect on the performance of the system. However, the most crucial test, the one that guarantees the usability of the interface, is the frames/second rate.

### 5.1 Frames Per Second

Frames/second is a measure of the frame rate of the system as the graphics are rendered. This test is considered successful if a frame rate of 20 frames per second (fps) or greater is recorded. 10 fps or fewer is considered an unsuccessful test, since research shows that as the frame rate drops to around 10 fps, the rendering becomes "choppy" and thus reduces user efficiency. Fortunately the results of the test showed that on the development machine, the frame rate measured well over 30 frames per second.

### 5.2 Problems Revealed Through Testing

Through the testing process, two significant problems were discovered:

The first of these problems dealt with the use of Direct3D. When the system was set to use Direct3D instead of OpenGL, the decals used to display the names of the subdirectories on the walls did not show up, thus limiting the usefulness of the system.

Secondly, when switching (ALT-TAB) between it and Microsoft Windows©, the 3D file management system crashed. This was due to compatibility issues with OpenGL in the Windows© environment.

## 6 Example of Use

The following sequence of steps describes an example of the "cut" operation:

Figure 2 shows a directory listing of the "new" directory, containing the subdirectories "new1", "new2", "new3", "new4", and eight "New Text Document".

Figure 3 displays one of those text documents being selected, using the 3D user interface. Note the two portals in the background for subdirectories "new3" and "new4".

Figure 4 and Figure 5 depict the selected text file, i.e. "New Text Document(2).txt", being removed from the directory with a "cut" operation.

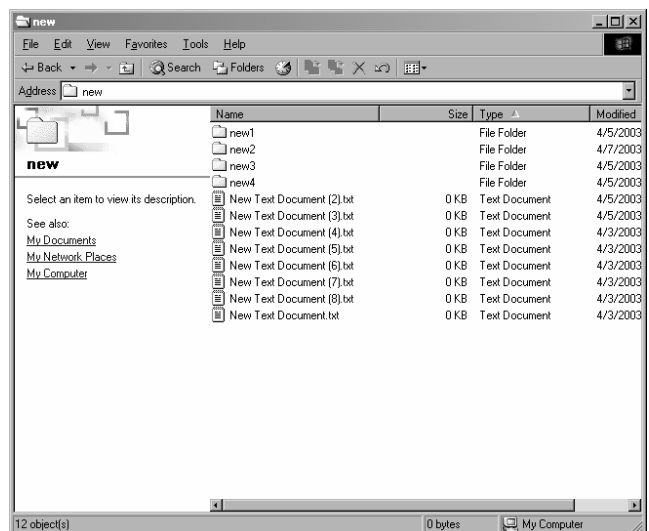

Figure 2: Directory listing of "new" directory.

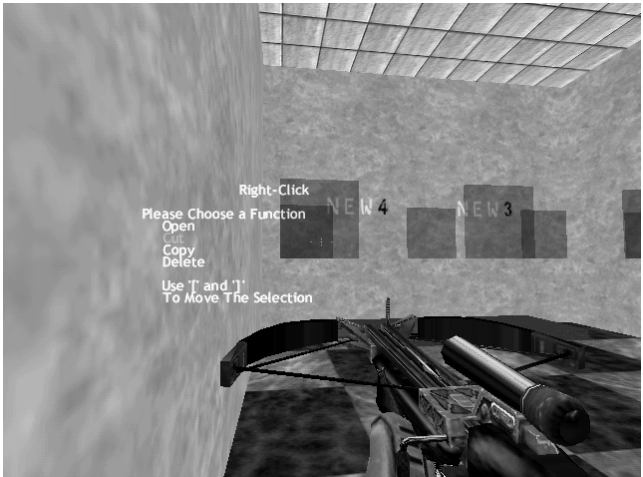

Figure 3: "cut" is selected from the right-click menu.

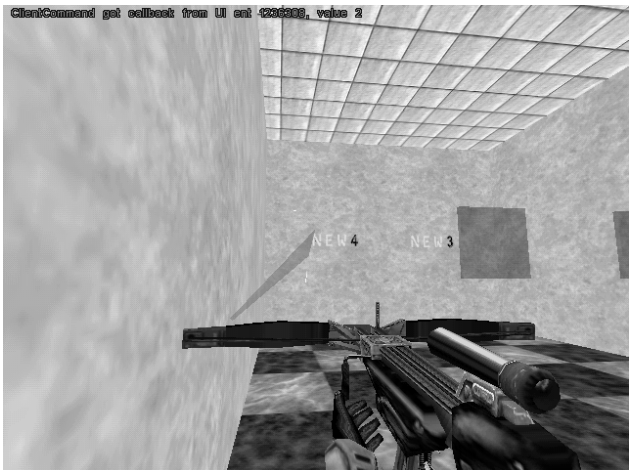

Figure 4: The selected file is "blown-up".

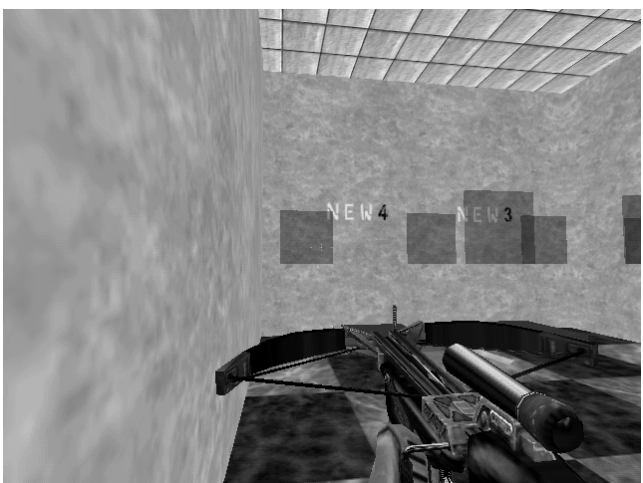

Figure 5: The selected file is removed.

# 7 Conclusions

3D interfaces offer a new way to interact with computers. It is very attractive to use a game-like interface for traditional manipulations of operating systems objects. It can seriously help in studying operating systems peculiarities. The process of creating the 3D file management system started with the identification of a gap within the computing industry; that is, a lack of 3D user interfaces incorporated into an operating system, even though the hardware to enable this exists.

Having defined this gap, it was necessary to come up with some important requirements for a system that would utilize a 3D GUI but would be based on a piece of software that was currently in use. After much thought it was decided that a file management system would be the best system to implement, as it contains features most users are aware of and it allows for a friendly 3D GUI to be developed for it.

The project has put forward a special user interface in which directories are represented as rectangular rooms. The path to parent/child directories are shown as transporters. Files are represented as panes of glass hovering in their appropriate room.

The design and implementation of the system originally started out as two separate activities plus the expansion of a 3D graphics engine. As the project progressed, these three tasks became intermingled and a very iterative process was followed.

Testing revealed the limitations of the software as it relates to the 3D engine used. Unfortunately the 3D engine chosen placed limitations on the maximum number of directories that could be rendered in the 3D user interface. However, this limitation leaves open the opportunity to develop a better engine to use as the backbone of the system.

The functionality of the file management system was accomplished by adding a menu class to the "client" source listing and adding the appropriate Windows API calls to existing methods. However, this came at a price. For the inexperienced user, the controls seem to be somewhat clumsy. It is recommended that, as the system matures, a 3D pointing device, such as a three-axis (X, Y, Z) mouse instead of the standard two axes, should be used.

As far as usability is concerned, there are still some unresolved issues with the interface. For instance, if a user needs to jump across several subdirectories, he/she must go back to the root directory and then proceed to the desired subdirectory. Additionally, the project would benefit if more user could try this new approach, so that comparisons could be made with currently metaphors. However, because there tremendous differences between a 2D metaphor and a 3D one, we anticipate that it is hard to come up with meaningful experiments to compare the two metaphors. If those trials are not suitable, we might end up comparing "apples" and "oranges".

Lastly, in order to make this sort of interface popular, we face three problems: First, within the operating systems domain, the use of 3D graphics is heavily under-used. Secondly, for ordinary users, on-the-fly creation of 3D "worlds" is a totally new trend, although it will soon be available to personal computer users. And thirdly, we still need empirical research to convince the average user of the benefits of having a 3D-GUI as a front-end from an operating system.